\documentclass[journal]{IEEEtran}

\usepackage{lineno,hyperref}
\usepackage{bm}
\usepackage{mathrsfs}
\usepackage{amssymb}
\usepackage{amsmath}
\usepackage{graphicx}
\usepackage{multirow}
\usepackage{textcomp,booktabs}
\usepackage[usenames,dvipsnames]{color}
\usepackage{colortbl}
\definecolor{mygray}{gray}{.9}
\usepackage{cite}
\usepackage[numbers,sort&compress]{natbib}
\modulolinenumbers[5]

\ifCLASSINFOpdf

\else

\fi

\begin{document}

\title{Recursive Geman-McClure Estimator for Implementing Second-Order Volterra Filter}

\author{Lu Lu, Wenyuan~Wang, Xiaomin Yang, Wei Wu and Guangya Zhu 	
	\thanks{Manuscript received July 25, 2018. This work was partially supported by the National Science Foundation of P.R. China under Grant 61701327.}
	\thanks{L. Lu, X. Yang and W. Wu are with the College of Electronics and Information Engineering, Sichuan University, Chengdu 610065, China. W. Wang is with School of Electrical Engineering, Southwest Jiaotong University, Chengdu, 610031, China. G. Zhu is with the College of Electrical and Information Technology, Sichuan University, Chengdu 610065, China. (e-mail: arielyang@scu.edu.cn (Corresponding author: Xiaomin Yang)).}}

\markboth{IEEE TRANSACTIONS ON CIRCUITS AND SYSTEMS-II: EXPRESS BRIEFS}
{Shell \MakeLowercase{\textit{et al.}}: Bare Demo of IEEEtran.cls for IEEE Journals}

\maketitle

\begin{abstract}
The second-order Volterra (SOV) filter is a powerful tool for modeling the nonlinear system. The Geman-McClure estimator, whose loss function is non-convex and has been proven to be a robust and efficient optimization criterion for learning system. In this paper, we present a SOV filter, named SOV recursive Geman-McClure, which is an adaptive recursive Volterra algorithm based on the Geman-McClure estimator. The mean stability and mean-square stability (steady-state excess mean square error (EMSE)) of the proposed algorithm is analyzed in detail. Simulation results support the analytical findings and show the improved performance of the proposed new SOV filter as compared with existing algorithms in both Gaussian and impulsive noise environments.
\end{abstract}


\IEEEpeerreviewmaketitle

\section{Introduction}
\IEEEPARstart{A}{daptive} Volterra filters have been thoroughly studied in diverse applications \cite{karakucs2017bayesian,mallouki2016analysis,sicuranza2004filtered,sicuranza2005multichannel}. However, such filters become computationally expensive when a large number of multidimensional coefficients are required. To overcome this problem, many methods were developed \cite{batista2016reduced,wang2016class,chen2016kernel,scarpiniti2018spline}. Among these, the second-order Volterra (SOV) filter was widely applied to identify nonlinear system with acceptable error level \cite{ogunfunmi2007}.

The impulsive noise is a great challenge for nonlinear system identification. It has been shown that the impulsive noise could be better modeled by $\alpha$-stable distribution \cite{shao1993signal}. A symmetric $\alpha$-stable distribution probability density function (PDF) is defined by means of its characteristic function \cite{shao1993signal} $\psi(\omega) = \exp \left\{-\gamma|\omega|^{\alpha}\right\}$, where $0 < \alpha \le 2$ is the \emph{characteristic exponent}, controlling the heaviness of the PDF tails, and $\gamma>0$ is the \emph{dispersion}, which plays an analogous role to the variance. Such $\alpha$-stable noise tends to produce ``outliers''. The recursive least square (RLS), based on the second-order moment, is not robust against outliers \cite{lu2016improved}. To address stability problem in impulsive noise environments, several RLS-based algorithms were proposed \cite{singh2010closed,navia2012combination,zou2001robust}. In \cite{navia2012combination}, a recursive least $p$-norm (RL$p$N) algorithm was proposed. However, this algorithm only achieves improved performance when $p$ closes to $\alpha$, where $p$ is the order of moments \cite{navia2012combination}. Another strategy is named as recursive least M-estimate (RLM) algorithm  which exploits the M-estimate function to suppress the outliers \cite{zou2001robust}. Although it is superior to several existing outlier-resistant methods, it suffers from performance degradation in highly impulsive noise environments.

Motivated by these considerations, we employ another M-estimator, named Geman-McClure estimate \cite{li2016learning}, for nonlinear system identification. Like the $\mathcal L_p$ estimator, the Geman-McClure estimator is a non-convex M-estimator, which is more efficient for learning system \cite{mandanas2017robust}. To the best of our knowledge, no adaptive algorithms can achieve improved performance in both Gaussian and $\alpha$-stable noise environments. By integrating the Geman-McClure estimator in the SOV filter structure, the proposed SOV recursive Geman-McClure algorithm achieves smaller steady-state kernel error as compared with the state-of-art algorithms. Note that a significant reduction is in fact vital from an engineering application perspective. Such a gain would likely justify the increase in computational complexity required to run the proposed algorithm. The fact that a considerable amount of mathematics is needed to derive the algorithms is of no consequence for practical applications, where the cost of hardware principally matters. In this paper, by proper application of mathematical concepts (even if at times cumbersome), we showed that key information about the signal environment could be extracted can be from their observation. Particularly, our main contributions are listed as follows: 

\noindent(i) The Geman-McClure estimator is first applied in SOV filter for the improved performance in the presence of $\alpha$-stable and Gaussian noises. 

\noindent(ii) The steady-state behaviour of the proposed algorithm is analyzed. 

\noindent(iii) We validate the theoretical findings and effectiveness of the proposed algorithm through simulations.

\vspace{-3mm}
\section{Problem Formulation}
\begin{figure}[!htb]
	\centering
	\includegraphics[scale=0.65] {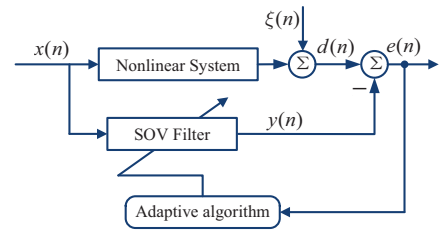}
	\caption{\label{1} Diagram of nonlinear system identification using SOV filter.}
	\label{Fig01}
\end{figure}
Fig. \ref{Fig01} shows the nonlinear system identification model based on the SOV filter, where $x(n)$ and $y(n)$ denote the input and output data, $e(n)=d(n)-y(n)$ denotes the error signal, $d(n)$ is the desired signal, and $\xi(n)$ is the noise signal. Given desired signal $d(n)$ that satisfies a model of the form
\begin{equation}
d(n) = \bm h_o^{\mathrm T}\bm x(n)+\xi(n)
\label{001}
\end{equation}
we want to estimate an $L \times 1$ unknown vector $\bm h_o$, where $L$ denotes the length of the SOV filter. The expanded input vector $\bm x(n)$ and the corresponding expanded coefficient vector ${\bm {\hat h}}(n)$ of the SOV system at time $n$ are expressed as
\begin{equation}
\begin{aligned}
\bm x(n)=&[x(n),x(n-1),\ldots,x(n-M+1), \\
& x^2(n),x(n)x(n-1),\ldots,x^2(n-M+1)]^{\mathrm T}
\label{002}
\end{aligned}
\end{equation}
\vspace{-3mm}
\begin{equation}
\begin{aligned}
{\bm {\hat h}}(n)=[h_1(0),\ldots,h_1(M-1),h_2(0,0),\ldots\\
,h_2(M-1,M-1)]^{\mathrm T}
\label{003}
\end{aligned}
\end{equation}
where $M$ denotes the length of the linear kernel, and $h_r$ stands for the $r$th-order Volterra kernel. Thus, the output of the SOV filter is expressed as
\begin{equation}
\begin{array}{l}
y(n) = {\bm {\hat h}}^{\mathrm T}(n)\bm x(n) = \sum\limits_{m_1=0}^{M-1} {h_1(m_1)} x(n - m_1) \\ 
+ \sum\limits_{m_1=0}^{M-1} {\sum\limits_{m_2=m_1}^{M-1} {h_2(m_1,m_2)} x(n-m_2)x(n - m_1)}.  
\end{array}
\label{004}
\end{equation}
In this case, $L=M(M+3)/2$. In practice, the noise signal $\xi(n)$ may be either Gaussian or non-Gaussian. Hence, it is very clear that efforts make sense in pursuing a more effective SOV-algorithm that satisfies faster convergence and smaller misalignment.
\vspace{-5mm}
\section{Proposed algorithm}

\begin{figure}[!htb]
	\centering
	\includegraphics[scale=0.4] {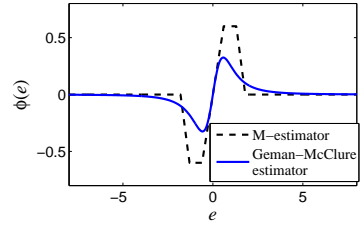}
	\caption{\label{2} Score functions of Hampel`s three-part redescending M-estimator and Geman-McClure estimator ($\sigma$=1, and three threshold parameters in M-estimate are set to 0.6, 1.3 and 1.8).}
	\label{Fig02}
\end{figure}

The conventional Geman-McClure estimator has the following form \cite{li2016learning}:
\begin{equation}
\Theta(e) = \frac {e^2}{\sigma^2+e^2} 
\label{005}
\end{equation}
where $\sigma$ is a constant that modulates the shape of the loss function. Fig. \ref{Fig02} shows the score functions $\phi(e)$ of the M-estimator and the Geman-McClure estimator, where $\phi(e)=\partial\Theta(e)/\partial e$. It can be observed that for larger values of $e$, the weight updation is small and thus the algorithm is stable in the presence of impulsive noise. For performance improvement, the type of recursive algorithms is usually preferred. Inspired by these merits, the cost function of the proposed algorithm is defined as follows: 
\begin{equation}
\begin{array}{rcl}
J(n) \triangleq \sum\limits_{k=1}^n {\lambda^{n-k}\frac{e^2(k,n)}{\sigma^2+e^2(k,n)}}  
\end{array}
\label{006}
\end{equation}
where $0\ll\lambda<1$ is the forgetting factor. The error signal $e(k,n)$ can be expressed as
\begin{equation}
e(k,n) = d(k) - \bm x^{\mathrm T}(k)\bm {\hat h}(n).
\label{007}
\end{equation}
Taking the gradient of $J(n)$ with respect to the weight vector ${\bm {\hat h}}(n)$, and letting the equation be zero, we have
\begin{equation}
\begin{array}{l}
\sum\limits_{k=1}^n \lambda^{n-k}\rho(k,n)\bm x(k){\bm x^{\mathrm T}}(k){\bm {\hat h}}(n) = \sum\limits_{k=1}^n {\lambda^{n-k}\rho(k,n)\bm x(k)d(k)}
\end{array}
\label{008}
\end{equation}
where $\rho(k,n)=\frac{\sigma^2}{(\sigma^2+e^2(k,n))^2}$ is the weighting factor. Then, the expression of ${\bm {\hat h}}(n)$ can be rewritten as:
\begin{equation}
\begin{array}{rcl}
\begin{array}{l}
{\bm {\hat h}}(n) = {\bm P}(n){\bm \theta}(n) = {\bm \Phi}^{-1}(n){\bm \theta}(n) 
\end{array}
\end{array}
\label{009}
\end{equation}
where ${\bm P}(n) = {\bm \Phi}^{-1}(n)$, ${\bm \Phi}(n) = \sum\limits_{k = 1}^n \lambda^{n-k}\rho(k,n)\bm x(k){\bm x^{\mathrm T}}(i)$
and ${\bm \theta}(n) = \sum\limits_{k = 1}^n \lambda^{n-i}\rho(k,n)d(k)\bm x(k)$. If $\rho(k,n) = 1$, the above update equation becomes the conventional RLS algorithm. If $\rho(k,n)\ne1$, $\bm P(n)$ and ${\bm \theta}(n)$ are the weighted autocorrelation matrix and the weighted cross-correlation vector of the optimal weights via $\rho(k,n)$. We have to recalculate (\ref{009}) at each iteration. In our previous studies, an online recursive method is considered to overcome this limitation \cite{lu2016improved}. By using this approach, $\bm P(n)$ and ${\bm \theta}(n)$ can be adapted by
\begin{equation}
\begin{array}{l}
\bm \Phi(n)\approx \sum\limits_{k = 1}^n \lambda^{n-k}\rho(k,k)\bm x(k){\bm x^{\mathrm T}}(k) \\
\;\;\;\;\;\;\;\;\;= \lambda{\bm \Phi}(n-1) + \rho(n,n)\bm x(n){\bm x^{\mathrm T}}(n),\\
\label{010}
\end{array}
\end{equation}
\vspace{-6mm}
\begin{equation}
\begin{array}{l}
{\bm \theta}(n) \approx \sum\limits_{k = 1}^n \lambda^{n-k}\rho(k,k)d(k)\bm x(k) \\
\;\;\;\;\;\;\;\;= \lambda {\bm \theta}(n-1) + \rho(n,n)\bm x(n)d(n).
\label{011}
\end{array}
\end{equation}

Using the matrix inversion lemma \cite{sayed2003fundamentals}, the adaptation equation of ${\bm P}(n)$ can be obtained as
\begin{equation}
{\bm P}(n) = \lambda^{-1}{\bm P}(n-1) - \lambda^{-1}{\bm \Psi}(n){\bm x^{\mathrm T}}(n){\bm P}(n-1)
\label{012}
\end{equation}
where ${\bm P}(0) = \zeta^{-1}{\bf I}$, ${\bf I}$ is an identity matrix, $\zeta=0.01$ is a small positive number, and the gain vector is
\begin{equation}
{\bm \Psi}(n) = \frac{\rho(n,n){\bm P}(n-1)\bm x(n)}{\lambda + \rho(n,n){\bm x^{\mathrm T}}(n){\bm P}(n-1)\bm x(n)}.
\label{013}
\end{equation}
Rewriting (\ref{009}) in a recursive way, we can obtain the following update equation for ${\bm {\hat h}}(n)$ 
\begin{equation}
{\bm {\hat h}}(n) = {\bm {\hat h}}(n-1) + {\bm \Psi}(n)[d(n) - {\bm x^{\mathrm T}}(n){\bm {\hat h}}(n-1)].
\label{014}
\end{equation}

$\emph{Remark 1}$: In the expression (\ref{014}), one can see an implicit relationship between ${\bm {\hat h}}(n)$ and $\rho(n,n)$. The algorithm requires an iterative approximation to the optimal solution, where $\rho(n,n)$ is calculated by using  ${\bm {\hat h}}(n-1)$, and the new value for ${\bm {\hat h}}(n)$ is obtained with the compute the value of $\rho(n,n)$.

$\emph{Remark 2}$: The proposed algorithm is easy to implement in the SOV filter, since it does not require any priori information on the noise characteristics. It requires two parameters ($\sigma$ and $\lambda$) to improve the overall nonlinear filtering performance.
\vspace{-5mm}
\section{Performance analysis}

In this section, given the Assumptions, we theoretically study the performance of the proposed SOV algorithm. Because the output of the Volterra-based algorithms linearly depend on the coefficients of the filter itself, we can follow the approach in \cite{sayed2003fundamentals,lee1993fast} for analyzing the mean performance and steady-state excess mean square error (EMSE) of SOV recursive Gemen-McClure algorithm.

To begin with, we define the weight deviation vector as $\widetilde{{\bm h}}(n) \triangleq {\bm h_o} - {\bm {\hat h}}(n)$. In steady-state, the \emph{a priori} error $e_a(n)$ and the \emph{a posterior} error $e_p(n)$ are defined as $e_a(n) \triangleq {\bm x^{\mathrm T}}(n)\widetilde{{\bm h}}(n-1),\;e_p(n) \triangleq {\bm x^{\mathrm T}}(n)\widetilde{{\bm h}}(n)$. The mathematical analysis needs the following assumptions.\\
\noindent \textbullet\; \emph{Assumptions}\\
a) The input signal $\bm x(n)$ is independent and identically distributed (i.i.d.) with zero-mean, and $\left \lVert\bm x(n)\right \rVert_{\bm \Omega}^2$ is approximately independent of the \emph{a priori} error $e_a(n)$ at steady-state, where $\left \lVert \bm x\right \rVert_{\bm \Omega}^2={\bm x}^{\mathrm T}{\bm \Omega}{\bm x}$ stands for the squared-weighted Euclidean norm of a vector.

\noindent b) The noise signal $\xi(n)$ is i.i.d. with zero-mean and variance $\sigma_\xi^2$.

\noindent c) $\xi(n)$ and $\bm x(n)$ are mutually independent.
\vspace{-3.5mm}
\subsection{Mean stability}

We can use $\widetilde{{\bm h}}(n)$ to rewrite the adaptation equation (\ref{014}) as
\begin{equation}
\begin{array}{rcl}
&\widetilde{{\bm h}}(n) = \widetilde{{\bm h}}(n-1)-\frac{\rho(n,n){\bm P}(n-1)\bm x(n)}{\lambda+\rho(n,n){\bm x^{\mathrm T}}(n){\bm P}(n-1)\bm x(n)} \\
&\bm\cdot [{d(n) - {\bm x^{\mathrm T}}(n){\bm {\hat h}}(n-1)}].
\label{017}
\end{array}
\end{equation}
Based on the definition of (\ref{012}), and using the matrix inversion lemma, we obtain the adaptaion of ${{\bm P}^{-1}}(n)$, which is ${{\bm P}^{-1}}(n) = \lambda^{n+1}\rho{\bf I} + \sum\limits_{k=0}^n \lambda^{n-k}\rho(k,k)\bm x(k)\bm x^{\mathrm T}(k)$. Then, we rewrite (\ref{017}) as
\begin{equation}
\begin{array}{l}
\widetilde{{\bm h}}(n) = \widetilde{{\bm h}}(n-1) - \mu(n){\bm P}(n-1){\bm x^{\mathrm T}}(n)e(n)
\label{019}
\end{array}
\end{equation}
where $\mu(n)=\frac{1}{\frac{\lambda}{\rho(n,n)}+{\bm x^{\mathrm T}}(n){\bm P}(n-1)\bm x(n)}$ and $e(n)=\bm x^{\mathrm T}(n)\widetilde{{\bm h}}(n-1)+\xi(n)$. Taking expectations of both sides of (\ref{019}) and using $e(n) \approx \bm x^{\mathrm T}(n)\widetilde{{\bm h}}(n-1)$ during the transient state, yields
\begin{equation}
\hskip-1em
\begin{array}{l}
{\mathrm E} \{\widetilde{{\bm h}}(n)\} = {\mathrm E} \{\widetilde{{\bm h}}(n-1)\} - {\mathrm E} \{\mu(n){\bm P}(n-1){\bm x^{\mathrm T}}(n)e(n)\}\\
\approx  {\mathrm E} \{\widetilde{{\bm h}}(n-1)\} - {\mathrm E} \left\{\mu(n){\bm P}(n-1){\bm x^{\mathrm T}}(n)\bm x(n)\right\} {\mathrm E} \{\widetilde{{\bm h}}(n-1)\}
\label{019a}
\nonumber
\end{array}
\end{equation}
where ${\mathrm E}\{\cdot\}$ denotes the statistical expectation operator. Therefore, the proposed algorithm can converge in the sense of mean if and only if
\begin{equation}
\begin{aligned}
0 < \lambda_{\max} \left({\mathrm E}\left\{\mu(n){\bm P}(n-1){\bm x^{\mathrm T}}(n)\bm x(n)\right\}\right) <2
\label{019b}
\end{aligned}
\end{equation}
where $\lambda_{\max}(\cdot)$ is the largest eigenvalue of a matrix. Based on the fact that $\lambda_{\max}({\bf A}{\bf B})< {\mathrm {Tr}}({\bf A}{\bf B})$ in (\ref{019b}) \footnote{$\lambda_{\max}({\bf A}{\bf B}) < {\mathrm {Tr}}({\bf A}{\bf B})$ is simply not true in the general case. However, if the input signal is persistently exciting, $\bm P(n)>0$ for all infinite $n$ and the matrix $\bm x(n)\bm x^{\mathrm T}(n)$ is nonnegative definite in (\ref{019b}). Hence, we have such inequality.}, we obtain
\begin{equation}
\begin{array}{l}
\lambda_{\max} \left({\mathrm E}\left\{ \frac{{\bm P}(n-1){\bm x^{\mathrm T}}(n)\bm x(n)}{\frac{\lambda}{\rho(n,n)}+{\bm x^{\mathrm T}}(n){\bm P}(n-1)\bm x(n)} \right\} \right) \\
<{\mathrm E}\left\{ \frac{{\mathrm {Tr}}({\bm x^{\mathrm T}}(n){\bm P}(n-1) \bm x(n))}{\frac{\lambda}{\rho(n,n)}+{\bm x^{\mathrm T}}(n){\bm P}(n-1)\bm x(n)} \right\} <1.
\label{019c}
\end{array}
\end{equation}
Consequently, the mean error weight vector of the proposed algorithm is convergent if the input signal is persistently exciting \cite{sayed2003fundamentals}.

\subsection{Mean-square stability}

Multiplying both sides of (\ref{019}) by $\bm x(n)$, we have the relationship between the \emph{a priori} and the \emph{a posteriori} errors, as below
\begin{equation}
\begin{array}{l}
e_p(n) = e_a(n) - \left \lVert \bm x(n)\right \rVert_{{\bm P}(n-1)\mu(n)}^2e(n){\color{red}.}
\label{020}
\end{array}
\end{equation}
Using the established energy conservation argument, the proposed algorithm can be expressed as
\begin{equation}
\begin{array}{l}
\widetilde{\bm h}(n) + \frac{{\mu(n)\bm P(n-1){\bm x^{\mathrm T}}(n)}}{\left \lVert \bm x(n)\right \rVert_{{\bm P}(n-1)\mu(n)}^2}e_a(n) = \\
{\quad\quad\quad\quad\quad\quad\quad}\widetilde{\bm h}(n-1) + \frac{\mu(n)\bm P(n-1){\bm x^{\mathrm T}}(n)}{\left \lVert \bm x(n)\right \rVert_{\bm P(n-1)\mu(n)}^2}e_p(n).
\label{021}
\end{array}
\end{equation}
Combining (\ref{019}) and (\ref{020}) and employing $\mu^{-1}(n){\bm P}^{-1}(n-1)$ as a weighting matrix for the squared-weighted Euclidean norm of a vector, we can get
\begin{equation}
\begin{array}{l}
|| \widetilde{{\bm h}}(n) ||_{{\mu^{-1}}(n){{\bm P}^{-1}}(n-1)}^2 + \frac{e^2_a(n)}{\left \lVert \bm x(n)\right \rVert_{\mu(n){\bm P}(n-1)}^2} \\
= || \widetilde{{\bm h}}(n-1) ||_{\mu^{-1}(n){{\bm P}^{-1}}(n-1)}^2 + \frac{e^2_p(n)}{\left \lVert\bm x(n)\right \rVert_{\mu(n){\bm P}(n-1)}^2}.
\label{022}
\end{array}
\end{equation}
In the SOV filter with the recursive Geman-McClure algorithm, the adaptive filter will converge to the optimum (minimum) EMSE at steady{\color{red}-}state. Therefore, we have
\begin{equation}
\begin{array}{l}
{\mathrm E}\left\{{||\widetilde{{\bm h}}(n)||_{\mu^{-1}(n){\bm P}^{-1}(n-1)}^2} \right\}\\
\approx {\mathrm E}\left\{{||\widetilde{{\bm h}}(n-1)||_{{\mu^{-1}}(n){\bm P}^{-1}(n-1)}^2}\right\}
\label{023}
\end{array}
\end{equation}
Taking expectations of both sides of (\ref{022}), and substituting (\ref{023}) into (\ref{022} yields
\begin{equation}
\begin{array}{l}
{\mathrm E}\left\{{\frac{e^2_a(n)}{\left \lVert \bm x(n)\right \rVert_{\mu(n){\bm P}(n-1)}^2}} \right\} = {\mathrm E}\left\{{\frac{e^2_p(n)}{\left \lVert \bm x(n)\right \rVert_{\mu(n){\bm P}(n-1)}^2}} \right\}.
\label{024}
\end{array}
\end{equation}
Substituting (\ref{020}) into (\ref{024}) results in
\begin{equation}
\begin{array}{l}
{\mathrm E}\left\{{\frac{e^2_a(n)}{\left \lVert \bm x(n)\right \rVert_{\mu(n){\bm P}(n-1)}^2}} \right\}
={\mathrm E}\left\{{\frac{e^2_a(n)}{\left \lVert \bm x(n)\right \rVert_{\mu(n){\bm P}(n-1)}^2}} \right\} \\ + {\mathrm E}\left\{{\left \lVert \bm x(n)\right \rVert_{\mu(n){\bm P}(n-1)}^2e^2(n)}\right\} - 2 {{\mathrm E}\left\{{e_a(n)e(n)} \right\}}.
\label{025}
\end{array}
\end{equation}
Therefore, in the steady{\color{red}-}state ($n\to\infty$), (\ref{025}) can be reduced to
\begin{equation}
\begin{array}{l}
{\mathrm E}\left\{{\left \lVert \bm x(n)\right \rVert_{\mu(\infty){\bm P}(\infty)}^2e^2(\infty)}\right\} = 2 {{\mathrm E}\left\{{e_a(\infty)e(\infty)} \right\}}.
\label{026}
\end{array}
\end{equation}
Considering Assumption c) and using $e(n)=e_a(n)+\xi(n)$, the left side of (\ref{026}) can be expressed as
\begin{equation}
\begin{aligned}
&{\mathrm E}\left\{{\left \lVert \bm x(n)\right \rVert_{\mu(\infty){\bm P}(\infty)}^2(e_a(\infty)+\xi(\infty))^2}\right\}\\
=&\;{\mathrm E}\left\{{\left \lVert \bm x(n)\right \rVert_{\mu(\infty){\bm P}(\infty)}^2e_a^2(\infty)}\right\}
+{\mathrm E}\left\{{\left \lVert \bm x(n)\right \rVert_{\mu(\infty){\bm P}(\infty)}^2\xi^2(\infty)}\right\}\\
&+2{\mathrm E}\left\{{\left \lVert \bm x(n)\right \rVert_{\mu(\infty){\bm P}(\infty)}^2e_a(\infty)\xi(\infty)}\right\}.
\label{027}
\end{aligned}
\end{equation}
According to Assumption a), (\ref{027}) is reduced to
\begin{equation}
\begin{array}{l}
\sigma_v^2{\mathrm E}\left\{ {\left \lVert\bm x(n)\right \rVert_{\mu(\infty){\bm P}(\infty)}^2} \right\} + {\mathrm E}\left\{ \left \lVert\bm x(n)\right \rVert_{\mu(\infty){\bm P}(\infty)}^2\right\}
 {\mathrm E}\left\{e_a^2(\infty)\right\}.   
\label{028}
\end{array}
\end{equation}
Reusing $e(n)=e_a(n)+\xi(n)$ and considering Assumption b), the right side of (\ref{026}) can be expressed as
\begin{equation}
\begin{array}{c}
2{{\mathrm E}\left\{{e_a(\infty)e(\infty)} \right\}} \approx 2{\mathrm E}\left\{{e^2_a(\infty)}\right\}.
\label{029}
\end{array}
\end{equation}
Using (\ref{028}) and (\ref{029}), it can be shown that
\begin{equation}
\begin{array}{l}
\sigma_\xi^2{\mathrm E}\left\{{\left \lVert\bm x(n)\right \rVert_{\mu(\infty){\bm P}(\infty)}^2} \right\} + {\mathrm E}\left\{{\left \lVert \bm x(n)\right \rVert_{\mu(\infty){\bm P}(\infty)}^2}\right\}\\
\bm\cdot {\mathrm E}\left\{e_a^2(\infty)\right\} = 2{\mathrm E}\left\{e_a^2(\infty) \right\}.
\label{030}
\end{array}
\end{equation}

Eq. (\ref{030}) can be rewritten as
\begin{equation}
\tau = \frac{\vartheta\sigma_\xi^2}{2 - \vartheta}
\label{031}
\end{equation}
where $\tau = {\mathrm E}\left\{{e_a^2(\infty)}\right\}$ and $\vartheta = {\mathrm E}\left\{{\left \lVert \bm x(n)\right \rVert_{\mu(\infty){\bm P}(\infty)}^2} \right\}
= {\mathrm E}\left\{{\frac{{\sigma^2{\mathrm {Tr}}(\bm x(n){\bm x^{\mathrm T}}(n){\bm{P}}(\infty))}}{{{({\sigma^2} + e^2(\infty))}^2}\lambda + \sigma^2{\mathrm {Tr}}(\bm x(n){\bm x^{\mathrm T}}(n){\bm P}(\infty))}} \right\}$.
where ${\mathrm {Tr}}(\cdot)$ is trace operation. Now, let us define the steady{\color{red}-}state mean value of ${{\bm P}^{-1}}(n)$ as 
${\bm {\mathcal P}^{-1}} \buildrel \Delta \over = \mathop {\lim}\limits_{n \to \infty} {\mathrm E}\left\{{\bm P}^{-1}(n)\right\} = \frac{{{\mathrm E}\{\rho(n,n)\} {\bm \Xi}(n)}}{1-\lambda}$ where ${\bm \Xi}(n)$ is the covariance matrix ${\bm \Xi}(n) = {\mathrm E}\{\bm x(n){\bm x^{\mathrm T}}(n)\}$. When the algorithm is close to the optimal EMSE. In this case, ${\mathrm E}\left\{{{\bm P}(\infty)} \right\}$ can be approximated as
\begin{equation}
\begin{array}{l}
{\mathrm E}\left\{{{\bm P}(\infty)} \right\} \approx {\left( {{\mathrm E}\left\{{{\bm P}^{-1}(\infty)}\right\}} \right)^{-1}} = \bm {\mathcal P} = \frac{(1 - \lambda){\bm \Xi}^{-1}(\infty)}{{\mathrm E}\left\{ {\frac{\sigma^2}{(\sigma^2+e^2(\infty))^2}} \right\}} \\
\approx \frac{{\mathrm E}\{{(\sigma^2 + e^2(\infty))}^2\} (1 - \lambda){\bm \Xi}^{-1}(\infty)}{\sigma^2}.
\label{034}
\end{array}
\end{equation}
For $0\ll\lambda<1$, we have $|e_a(n)| \ll |\xi(n)|$ at steady-state. Finally, substituting (\ref{034}) and into (\ref{031}), we arrive at
\begin{equation}
\epsilon = \frac{\sigma_\xi^2(1-\lambda)L \varphi}{2-(1-\lambda)L \varphi}
\label{035}
\end{equation}
where $\varphi = {\mathrm E}\left[{\frac{{\mathrm E}\left[\left(\sigma^2+\xi^2(n)\right)^2\right]}{\lambda {{(\sigma^2+\xi^2(n))}^2} + {\mathrm E}\left[\left(\sigma^2+\xi^2(n)\right)^2\right](1-\lambda)L}} \right]$.
Note that it is very hard to further simplify (\ref{035}). The theoretical result contains a random variable $\xi(n)$, but after the expect operation, we can obtain an exact value. Furthermore, (\ref{035}) is also applicable to the analysis of linear recursive Geman-McClure algorithm.

Finally, we compare the computational complexities of the algorithms, as shown in Table \ref{Table01}, where $N_w$ denotes the length of sliding-window in the SOV-RLM algorithm. The arithmetic complexity of the proposed algorithm is comparable to that of the SOV-RLS algorithm, except for the $L+3$ more multiplications, 1 addition and 1 division in (\ref{013}). 

\begin{table}[tbp]
	\scriptsize
	\centering
	\caption{Summary of the computational complexity in each iteration.}
	\doublerulesep=0.5pt
	\begin{tabular}{p{1.2cm}|c c p{1.8cm}}
		\hline
		\hline
		\textbf {Algorithms} &\textbf {Multiplications} &\textbf {Additions} &\textbf {Other operations} \\ \hline 
		SOV-RLS &$2L^2+4L$ &$2L^2+2L$ &1 division \\ \hline
		SOV-RLM &$2L^2+5L$ &$2L^2+2L$ &1 division and ${\mathcal O}(N_w{\mathrm {log}}N_w)$ \\ \hline
		SOV-RL$p$N &$2L^2+5L+2$ &$2L^2+2L+1$ &2 divisions and $p$-power operation \\ \hline
		\rowcolor{mygray}
		Proposed algorithm &$2L^2+5L+3$ &$2L^2+2L+1$ &2 divisions \\ 
		\hline 
		\hline
	\end{tabular}
	\label{Table01}
\end{table}

\section{Simulation results}

To verify the theoretical findings and to illustrate the effectiveness of the proposed algorithm, we present simulations when implemented in Matlab R2013a running on a 2.1GHz AMD processor with 4GB of RAM. The EMSE and normalized mean square deviation (NMSD) $= 20{\mathrm {log}}_{10} {\mathrm E}\{{|| {\bm {\hat h}}(n)-\bm h_o ||_2}\}/{||\bm h_o||_2}$ are employed to evaluate the performance. The results are averaged over 300 independent simulations. 
\subsection{Gaussian scenarios}
\begin{figure}[!htb]
	\centering
	\includegraphics[scale=0.4] {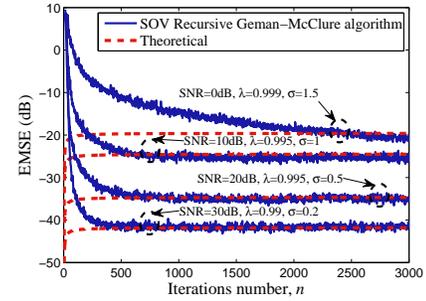}
	\caption{\label{3} Theoretical and simulation EMSEs for proposed algorithm.}
	\label{Fig03}
\end{figure}

\begin{table} [tbp]
	\scriptsize
	\centering
	\caption{EMSEs for the proposed algorithm ($\lambda$=0.99).}
	\doublerulesep=0.4pt
	\begin{tabular}{ccc|cc}
		\hline
		\hline
		\multicolumn{1}{ c }{\multirow{2}{*} {\textbf{SNR}}} & \multicolumn{1}{c}{\multirow{2}{*} {$\sigma$}} & \multicolumn{1}{ c }{\multirow{2}{*} {$L$}} & \multicolumn{2}{|c}{\textbf{EMSE}} \\ \cline{4-5}
		&  & & \textbf {Theory}  & \textbf {Simulation} \\
		\hline
		25dB& 0.5 & 14 \cite{lu2016improved}& $-$36.66dB & $-$36.75dB \\
		40dB& 1.8 & 14 \cite{lu2016improved}& $-$51.64dB & $-$51.83dB \\
		20dB& 0.2 & 14 \cite{lee1993fast}& $-$32.27dB & $-$31.77dB \\
		30dB& 0.45 & 14 \cite{lee1993fast}& $-$41.70dB & $-$41.83dB \\
		10dB& 0.9 & 20 \cite{kalluri1999general}& $-$20.65dB & $-$20.14dB \\
		30dB& 0.6 & 20 \cite{kalluri1999general}& $-$39.83dB & $-$40.40dB \\
		\hline
		\hline
	\end{tabular}
	\label{Table02}
\end{table}
\begin{figure}[!htb]
	\centering
	\includegraphics[scale=0.43] {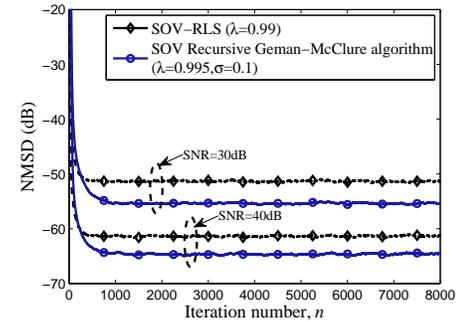}
	\caption{\label{4} NMSD curves of the algorithms for Gaussian environments.}
	\label{Fig04}
\end{figure}
In this example, we provide simulation verification for the analysis. The Gaussian distribution with zero-mean and unit variance is adopted to generate $\bm x(n)$ and $\xi(n)$. Fig. \ref{Fig03} plots the simulation and theoretical results of the EMSEs for different signal-to-noise ratio (SNR) and parameter settings. The unknown plant is a 14-tap nonlinear system, which is presented by \cite{lu2016improved}. It is observed that the simulations agree well with the analysis results in all scenarios. Next, we compare the theoretical and the simulation results of the steady-state EMSEs for different unknown systems. The input and noise signals used have similar characteristics as in Fig. \ref{Fig03}. The unknown parameter vector $\bm h_o$ has $L=14$ or $20$ entries and is defined by \cite{lee1993fast,kalluri1999general}. Table \ref{Table02} provides the simulation and theoretical EMSE values with different Volterra systems. This shows agreements in the SOV filter in spite of different SNRs, parameter settings and unknown systems. The difference between simulation results and theory arises from some approximations and assumptions used for deriving (\ref{035}).

{\color{red}W}e compare the convergence rate and steady-state kernel behaviour of the proposed algorithm with existing algorithms. The unknown system is a 14-tap nonlinear system \cite{lu2016improved}. A zero-mean white Gaussian noise (WGN) is used as the input signal. We observe that the proposed algorithm outperforms the standard SOV-RLS algorithm by nearly 5dB in steady-state where the noise signal is WGN with different SNRs (Fig. \ref{Fig04}).

\vspace{-5.5mm}
\subsection{Non-Gaussian scenarios}

\begin{table}[tbp]
	\scriptsize
	\centering
	\caption{Steady-state NMSDs of the proposed algorithms for different $\sigma$ with similar convergence rate ($\alpha=1.25, \gamma=1/15$).}
	\doublerulesep=0.1pt
	\begin{tabular}{c|p{0.67cm}<{\centering}|p{0.67cm}<{\centering}|p{0.67cm}<{\centering}|p{0.67cm}<{\centering}|p{0.67cm}<{\centering}|p{0.67cm}<{\centering}}
		\hline \hline
		$\sigma$                                                    & $0.05$ & $0.2$ & $0.3$ & $0.4$ & $1$ & $1.5$ \\ \hline
		\begin{tabular}[c]{@{}l@{}} \textbf{Steady-state}\\ \textbf{NMSD (dB)}\end{tabular} &$-44.24$ &$-43.00$ &$-45.53$ &$-41.19$ &$-38.76$ &$-37.57$ \\ \hline \hline
	\end{tabular}
	\label{Table03}
\end{table}
\begin{figure}[!htb]
	\centering
	\includegraphics[scale=0.43] {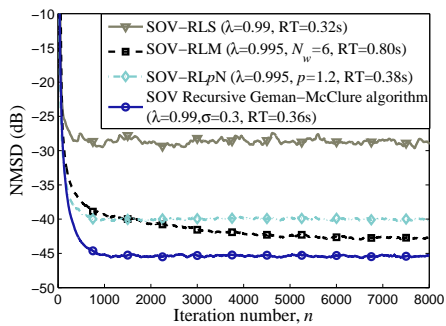}
	\caption{\label{6} NMSD curves of the algorithms in $\alpha$-stable noise ($\alpha=1.25, \gamma=1/15$).}
	\label{Fig06}
\end{figure}
In the second example, we repeat the same simulations as above, but this time using $\alpha$-stable noise as the noise signal. First, we study the effect of $\sigma$ on the performance of the SOV recursive Geman-McClure algorithm under the impulsive noise. For sake of fair comparison, we use the same fixed forgetting factor $\lambda=0.99$ to obtain a similar convergence rate. The results are shown in Table \ref{Table03}. It is indicated that the performance deteriorates quickly with the increasing of $\sigma$ when $\sigma$ is greater than 0.4. For an overall consideration of steady-state NMSD performance and convergence rate, the best choice is $\sigma=0.3$ in this example. Hence, we fix $\sigma=0.3$ in following simulations. Fig. \ref{Fig06} illustrates the NMSD performance of algorithms in the presence of $\alpha$-stable noise. One can also see that in impulsive noise scenario, the proposed SOV recursive Geman-McClure algorithm achieves better performance than its SOV-based counterparts \footnote{The SOV-RLM and SOV-RL$p$N algorithms can be derived by extending algorithms \cite{navia2012combination,zou2001robust} to the SOV filter structure.}. In the legend of the figure, we use the RT to denote the run time for algorithm. It can be seen that the run time of the proposed algorithm fall in between the RLS and RL$p$N. Since the proposed algorithm achieves improved performance in both cases, we can conclude that the SOV recursive Geman-McClure algorithm has robust performance for various scenarios and is an excellent alternative to the SOV-RLS algorithm for nonlinear system identification task.

\section{Conclusion}
In this paper, we proposed a recursive Geman-McClure algorithm based on SOV filter, which was derived by minimizing the Geman-McClure estimator of the error signal. Detailed steady-state analysis was presented. One of the advantages of the proposed algorithm is that it has only two parameters, the constant $\sigma$ and the forgetting factor $\lambda$, which have quite wide ranges for choice to achieve excellent performance. We carried out computer simulations that support the analytical findings and confirm the effectiveness of the proposed algorithm. Note that the variance of $\alpha$-stable noise is infinite, so it is very difficult, if not impossible, to conduct a steady-state performance analysis of the proposed algorithm. Similarly, theoretical analysis of global stability and convergence is also very hard and rare for the Volterra filters in the presence of $\alpha$-stable noise. Rigorous mathematical analysis is lacking in a long period. We leave these investigations for future work.

\ifCLASSOPTIONcaptionsoff
\newpage
\fi

\footnotesize
\bibliographystyle{IEEEtran}
\bibliography{IEEEabrv,mybibfile}

\end{document}